\documentclass[10pt,aps,prl,preprintnumbers,twocolumn,nofootinbib]{revtex4-1}
\pdfoutput=1
\usepackage[latin1]{inputenc}
\usepackage{amsmath}
\usepackage{amsfonts}
\usepackage{amssymb}
\usepackage{mathtools}
\usepackage{graphicx}
\usepackage{hyperref}
\usepackage{slashed}
\usepackage{tikz}

\usepackage[left=1.25cm,right=1.25cm,top=2cm,bottom=2cm]{geometry}
\usepackage{color}

\usepackage{ulem} 

\newcommand{\dd}{ \mathrm{d}}

\usetikzlibrary{decorations.pathmorphing}

\tikzset{snake it/.style={decorate, decoration=snake}}
\begin{document}

\preprint{CP3-Origins-2017-035 DNRF90}

\title{Neutron Star Stability in Light of the Neutron Decay Anomaly}

\date{\today}

\author{Benjam\'\i{}n Grinstein}
\email[E-mail: ]{bgrinstein@ucsd.edu}
\affiliation{Department of Physics, UC San Diego, La Jolla CA 92093, USA}
\author{Chris Kouvaris}
\email[E-mail: ]{kouvaris@cp3.sdu.dk}
\author{Niklas Gr\o nlund Nielsen}
\email[E-mail: ]{ngnielsen@cp3.sdu.dk}
\affiliation{CP${}^3$-Origins, University of Southern Denmark, Campusvej 55, DK-5230 Odense, Denmark}

\begin{abstract}
\noindent 
A recent proposal suggests that experimental discrepancies on the lifetime of  neutrons can be resolved if  neutrons decay to dark matter. At the same time it has been demonstrated that such a decay mode would soften the nuclear equation of state resulting in neutron stars with a maximum mass much below  currently observed ones.
In this paper we demonstrate that appropriate dark matter-baryon interactions can accommodate neutron stars with mass above 2 solar masses.  We also show that dark matter self-interactions could also help neutrons stars reach 2 solar masses provided that dark matter is of asymmetric nature.
\\[.1cm]
{\footnotesize  \it Preprint: CP3-Origins-2018-042 DNRF90}

\end{abstract}

\maketitle

\paragraph{\bf Introduction:} A puzzling discrepancy between the neutron lifetime as measured in bottle and beam experiments has persisted for more than 20 years~\cite{Patrignani:2016xqp}. In bottle experiments, a number of neutrons $N$ are cooled down and stored in a container. At later times the remaining number of neutrons is counted and fit to a decaying exponential $\propto \exp(-t/\tau_\text{bottle})$. 
In beam experiments, the rate of protons
  emitted from a beam of neutrons is  fit to $\dd N/\dd t = -N/\tau_\text{beam}$. The current PDG world average for the neutron lifetime in bottle experiments is $\tau_\text{bottle} = 879.6 \pm 0.6$~s, whereas the beam experiment measurements are significantly longer with a world average at $\tau_\text{beam} = 888.0 \pm 2.0$~s. The discrepancy is currently at the $4\sigma$ level.

Calculation of the neutron lifetime requires the nucleon axial coupling, $g_A$, as input.
Both a global fit \cite{Gardner:2013aya} and a recent 1\% lattice-QCD determination
\cite{Chang:2018uxx} give $\tau_n\approx 885$~s. However, if only $g_A$ determinations dating from
after 2002 are used, the resulting calculated lifetime agrees remarkably well with $\tau_{\text{bottle}}$ \cite{Czarnecki:2018okw}. 
A resolution involving new physics is not out of the question, and an exciting proposal by Fornal and Grinstein~\cite{Fornal:2018eol} recently sparked significant interest in the subject. They suggested an invisible decay of the neutron into dark matter (DM) in order to explain the neutron decay anomaly.
Since the beam measurements are only sensitive to beta decay, one would expect $\tau_\text{beam} = \tau_\text{bottle}/\text{Br}(n\to p+e+\bar{\nu}_e)$, suggesting that $\text{Br}(n\to p+e+\bar{\nu}_e) \approx 99\%$ with the remaining $1\%$ branching into DM.
 However, it was quickly realized that such an invisible decay channel of the neutron would lead to partial conversion of neutrons to DM inside a neutron star (NS). In such a case  the nuclear equation of state (EoS) softens so much at high densities that  makes it impossible for a NS to support masses above
 2$M_\odot$~\cite{McKeen:2018xwc,Baym:2018ljz,Motta:2018rxp}. This contradicts current observations setting  the maximum NS mass above $\sim$2$M_\odot$.  
 
 The authors of~\cite{Cline:2018ami} studied the NS stability including  DM self-interactions mediated by dark photons, concluding that although such interactions provide enough pressure to support  2$M_\odot$ NSs, the DM particle that the neutron decay to, must represent at best only a small fraction of the overall DM density of the Universe because annihilations of this DM type to the mediators in the early Universe cannot provide the right DM relic density. 
Models which induce DM self-interactions can alleviate tensions between numerical
simulations of collisionless cold DM (CCDM) and astrophysical observations.  
The ``core-cusp problem" refers to the fact that observations of
dwarf galaxies are consistent with flat density profiles in their central regions
\cite{Moore:1994yx,Flores:1994gz}, while N-body simulations of CCDM predict cuspy profiles
\cite{Navarro:1996gj}.  Another potential problem is that the number of satellite galaxies in the
Milkly Way is smaller than the number expected based again on simulations of CCDM
\cite{Klypin:1999uc,Moore:1999nt,Kauffmann:1993gv,Liu:2010tn,Tollerud:2011wt,Strigari:2011ps}.
Moreover, there is the  ``too big to fail" problem,  that
simulations predict dwarf galaxies in a mass range  not  observed until today, yet such dwarf
galaxies are too large not to have hosted stars by now
\cite{BoylanKolchin:2011de}. The solution of these problems is not currently known, and in principle
there can be different explanations for the various issues of the CCDM paradigm.    Simulations including DM self-interactions
suggest that they have the effect of smoothing out cuspy density profiles, and could solve the other
aforementioned problems of CCDM as well \cite{Dave:2000ar, Vogelsberger:2013eka, Rocha:2012jg}. These simulations prefer a self-interaction cross section of $0.1$~cm$^2$/g $\lesssim \sigma/m \lesssim 10$~cm$^2$/g.  There are, however, upper bounds on $\sigma/m$ from a number of sources, including the preservation of ellipticity of spiral galaxies \cite{Feng:2009mn,Feng:2009hw}.  The allowed parameter space from these constraints nonetheless intersects the range of cross sections which can resolve the small-scale issues of CCDM, in the range $0.1$ cm$^2$/g $\lesssim \sigma/m \lesssim 1$ cm$^2$/g.

In this paper, we study repulsive DM-baryon interactions, which allow to: i) accommodate the decay of neutrons to DM  and explain the neutron decay anomaly, ii) produce NSs with mass over 2$M_\odot$, and iii) can account for all of the DM relic density. Furthermore, the interactions we assume will induce DM self-interactions which may mitigate the CCDM problems we mentioned. Our solution to the neutron star stability problem is qualitatively different than those previously proposed using repulsive DM self-interactions~\cite{Cline:2018ami}.  Repulsive self-interactions stiffen the equation of state by raising the DM chemical potential and tuning down the DM to baryon fraction in equilibrium. Repulsive DM-baryon interactions can also stabilize the star, {not} by stiffening the EoS, but by making it energetically expensive to produce DM particles in a pure baryon medium (and vice versa). Therefore the cross interactions can impede the creation of any significant amount of DM inside the NS, thus maintaining NSs almost pure (without DM present) despite the fact that free neutrons could decay to DM. 
 
Before we present a specific example of a microscopic model that can
give all this, it is instructive to show how
DM-neutron interactions affect the stability of NSs. We model the
interaction by a vector or scalar boson mediated Yukawa potential
  \begin{equation}
    \label{eq:Yukawa} 
 U  = \pm \frac{g_\chi g_n}{4\pi} \frac{e^{-m_\phi r}}{r},
\end{equation}
where $+$ $(-)$ is for vector (scalar) exchange, $g_{\chi,n}$ are the couplings to
DM and neutron respectively and $m_\phi$ is the mass of the mediator. Couplings with equal
(opposite) sign result in repulsion (attraction) for vectors and vice versa for scalars. DM
self-interactions have a similar potential with coupling $g_{\chi}^2$
in place  of $g_{\chi}g_n$.
None of the  models proposed to date that explain the neutron decay anomaly lead to a repulsive cross
interaction, with the exception of Model 1 of~\cite{Fornal:2018eol} that has a photon in the
final state. However it couples to the neutron via a magnetic dipole interaction, which can be
attractive or repulsive depending on its orientation. We
  expect that neutrons will occupy equally spin up
and down states, and therefore such interactions will average to zero, hence not suitable in our case.

\begin{figure}
\includegraphics[width=0.45\textwidth]{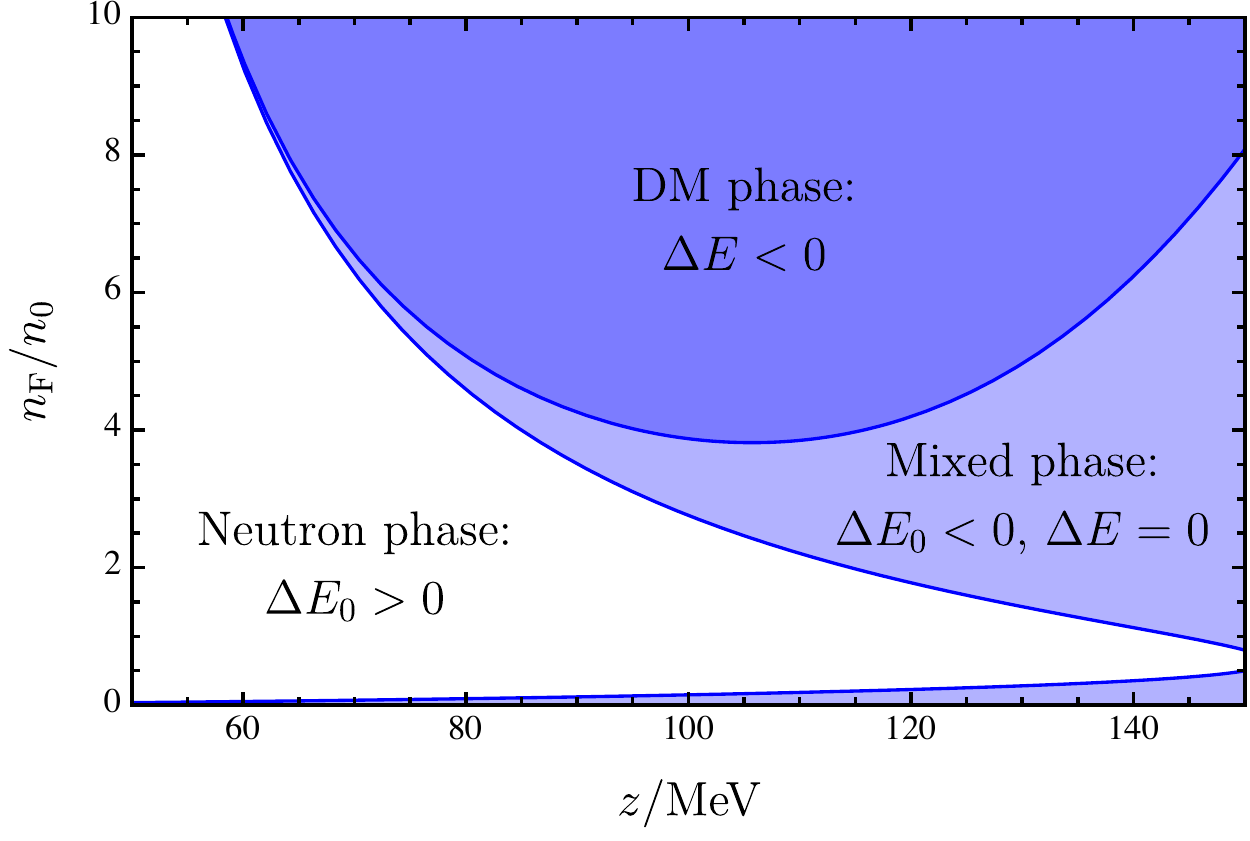}
\caption{The phase diagram using Eq.~\ref{Eq: free energy cost} and the SLy-4 EoS~\cite{Douchin:2001sv} assuming $m_n-m_\chi = 1$~MeV. In the white region there is an energy cost associated to creating any DM particles and pure neutron matter is preferred. In the light blue region, a chemical equilibrium exists with both neutron matter and DM. In the dark blue region, pure DM is thermodynamically preferred. For $z\lesssim 71$~MeV, fermion densities $<6n_0$ favor neutron matter and therefore do not affect NSs with mass smaller than $2M_\odot$.}
\label{Fig: ncrit}
\end{figure}

\paragraph{\bf Equation of state:} The energy density in a NS with DM and the above interaction is
\begin{equation}
\label{eos}
	\varepsilon(n_n,n_\chi)= \varepsilon_\text{nuc}(n_n)+\varepsilon_\chi(n_\chi)+\frac{n_\chi n_n}{2z^2},
\end{equation}
where $n_{n,\chi}$ are the neutron and DM number densities respectively, $\varepsilon_\text{nuc}$ is the energy density due to nuclear interactions,  $\varepsilon_\chi$ is the energy density of DM and the last term is the Yukawa energy density due to $n$-$\chi$ interactions in the mean field approximation where $z\equiv m_\phi/\sqrt{|g_\chi g_n|}$.
The last term in the energy density  is valid as long as the star is large compared to the Yukawa screening length, i.e. $R\gg 1/m_\phi$. Notice that the cross interaction depends simply on one parameter $z$. Long range forces are severely constrained and therefore we are going to assume that the mediator has a  mass. The DM energy density including DM self-interactions is
\begin{align}
\label{eps}
\varepsilon_\chi&= \frac{m_\chi^4}{8\pi^2}\left[x\sqrt{1+x^2}\left(1+2x^2\right)-\log\left(x+\sqrt{1+x^2}\right)\right]\pm\frac{n_\chi^2}{2z'^2},\notag \\
x&=\frac{\left(3\pi^2 n_\chi\right)^{1/3}}{m_\chi},
\end{align}
where $z'\equiv m_\phi/g_\chi$ and it is understood that the last term corresponds to DM self-interactions with $+$ ($-$) sign being repulsive (attractive).
The free energy cost at zero temperature associated to creating a DM particle at fixed total number density $n_\text{F} = n_n+n_\chi$ is just the change in internal energy, i.e.
\begin{equation}
\Delta E \equiv \frac{\partial \varepsilon(n_\text{F}-n_\chi,n_\chi)}{\partial n_\chi}=\mu_\chi(n_\chi)-\mu_\text{nuc}(n_n) + \frac{n_\text{F}-2n_\chi}{2z^2},\label{Eq: free energy cost}
\end{equation}
where $\mu_i$ ($i=\chi,\text{nuc}$) represent the chemical potentials of DM and neutrons respectively. A chemical equilibrium exists when $\Delta E=0$. In a pure neutron environment where no DM is present the energy cost is
\begin{equation}
\label{E0}
\Delta E_0= \left.\Delta E\right\vert_{n_\chi=0}=m_\chi-\mu_\text{nuc}(n_\text{F}) + \frac{n_\text{F}}{2z^2}.
\end{equation}
Notice, that the nuclear chemical potential $\mu_\text{nuc}>m_n$ and $m_n>m_\chi$ for the dark decay
to take place. Therefore in the absence of DM self-interactions, weak DM-neutron interactions (large
$z$) makes neutron conversion thermodynamically
favored~\cite{McKeen:2018xwc,Baym:2018ljz,Motta:2018rxp}, whereas stronger DM-neutron cross
interactions (small $z$) lead to a large energy cost for converting neutrons to DM that makes it
energetically favored to have zero DM density. For the nuclear EoS, we have chosen the
SLy-4~\cite{Douchin:2001sv} which is a nuclear EoS without a quark core and the power law EoS
$V_{3\pi}+V_\text{R}$~\cite{Gandolfi:2011xu} (which was also used in
Ref.~\cite{Cline:2018ami}). Both EoS we are using can in isolation support NSs with a mass larger
than $2M_{\odot}$, hence are consistent with observational data. As depicted in Fig.~\ref{Fig:
  ncrit}, for a given strength of DM-neutron cross interaction (i.e., for a given $z$),
there are three possibilities: i) $\Delta E_0>0$ i.e., the system is in a pure neutron phase simply because there is an energy cost to create DM. As can be seen by inspection of~Eq.(\ref{E0}), by strengthening cross interactions (i.e., by reducing $z$), the system can always enter the $\Delta E_0>0$ regime. This is also shown graphically in Fig.~\ref{Fig: ncrit}, where for a given total density $n_\text{F}$, there is always a $z$ below which $\Delta E_0>0$ and no DM particles are present.    ii) $\Delta E_0<0$ and $\Delta E=0$ for some value of $n_{\chi}<n_\text{F}$. In this case $\Delta E_0<0$  means that it is energetically favored to convert some of the neutrons to DM. The condition $\Delta E=0$ insures chemical equilibrium and by enforcing this condition using  Eq.~(\ref{Eq: free energy cost}), one can determine the amount of DM present (i.e., $n_\chi$). iii) The last regime satisfies $\Delta E<0$ for any $n_\chi$, which means that effectively all neutrons have converted to DM particles.
We measure number densities in units of the nuclear density $n_0= 0.16$~fm$^{-3}$. 
We will mainly focus on DM-baryon cross interaction as the mechanism allowing NSs to reach 2$M_{\odot}$. However, we will also comment on the case where DM self-interactions provide the needed support for heavy NSs.

In order to estimate how heavy NSs can be, we use Eq.~(\ref{eos}) with our two choices of nuclear EoS~\cite{Douchin:2001sv, Gandolfi:2011xu}. The pressure is derived from the relation $P = n_\text{F}^2 \dd (\varepsilon/n_\text{F})/\dd n_\text{F}$. In the limit of zero temperature (an excellent approximation for NSs), knowing the pressure and the energy density as a function of $n_{\chi}$, $n_{\text{nuc}}$ and the parameter $z$, allows us to solve the relativistic hydrostatic equilibrium described by the Tolmann-Oppenheimer-Volkoff  equation. By scanning the central fermion density, we find the maximum mass for the NS. Recall that  $n_{\chi}$ is uniquely fixed from the chemical equilibrium condition $\Delta E=0$ of Eq.~(\ref{Eq: free energy cost}). If $\Delta E_0>0$, we obviously have $n_\chi=0$. 
Therefore NSs can be as heavy as 2$M_{\odot}$ based on the fact that no DM is present to soften the EoS {(corresponding to a central density of $6n_0$ using the SLy-4 EoS shown in Fig.~\ref{Fig: ncrit})}. For our choices of EoS this corresponds to $z\lesssim 71$~MeV. No stable equilibrium exists if the central density is in the pure DM phase, and the heaviest stable configuration in the mixed phase is around $1.5M_\odot$.

\paragraph{\bf A microscopic model:} We move on to suggest a concrete model satisfying the two basic assumptions i.e., to provide the mechanism for the neutron decay to DM and provide repulsive cross interactions with $z\lesssim 71$~MeV. The potential in Eq.~\eqref{eq:Yukawa} can arise from scalar or vector exchange. In the former case the repulsive force requires opposite sign charges of $\chi$ and $n$, while in the latter same sign charges are required. 
In this example, we choose a scalar mediator, and realize the model as a  modification of model~2 of Ref.~\cite{Fornal:2018eol}. It requires four particles beyond the SM: a scalar $\Phi=(3,1)_{-1/3}$ (color triplet, weak singlet, hypercharge $-1/3$), two Dirac fermions  $\tilde\chi$ and $\chi$ (the DM particle), and a scalar  $\phi$ (all SM singlets). The neutron decay to $\chi+\phi$ is mediated by the very heavy $\Phi$ and the fermion $\tilde\chi$ with a mass that may be heavier than the neutron, through interactions given by 
\begin{align}
\mathcal{L} =&   \lambda_q \,\epsilon^{ijk}\, \overline{u^c_L}_{i}\, d_{Rj} \Phi_k + \lambda_\chi\Phi^{*i}\bar{\tilde\chi} \,d_{Ri}   +  \lambda_\phi  \,\bar{\tilde\chi}\, \chi \,\phi \notag\\
& + \mu  H^\dag H \phi + g_\chi\bar{\chi}\chi \, \phi+ {\rm h.c.} , 
\end{align}
where $d_R$ and $u_R$ are the standard model singlet quarks of charge $-1/3$ and $2/3$. The dark neutron decay takes place with interactions in the first line of the Lagrangian. The baryon numbers for the particles $\Phi$, $\tilde \chi$, $\chi$ and $\phi$ are chosen to be $-2/3$, $1$, $1$ and $0$, respectively. Compared to model 2 in Ref.~\cite{Fornal:2018eol} the baryon numbers of $\chi$ and $\phi$ are exchanged which allows additional terms in the Lagrangian, in particular a Higgs portal and a vertex with $\chi$ and $\phi$. The interaction through the Higgs portal induces an effective interaction with the neutron $g_n\bar{n}n \phi$ where
\begin{equation}
g_n = \frac{\mu \sigma_{\pi n}}{m_h^2},
\label{Eq: g_n}
\end{equation}
with $m_h = 125$~GeV being the Higgs mass and $\sigma_{\pi n} =  \sum_q \langle n|m_q \bar{q}q|n\rangle\approx 370$~MeV~\cite{Finkbeiner:2008qu}, where the sum runs over all quark flavors.
The model can incorporate the neutron decay to DM and DM relic density just as described in~\cite{Fornal:2018eol}. The repulsive DM-neutron interactions  require $g_ng_\chi<0$. To get the right interaction strength $z\lesssim 71$~MeV, we must consider constraints on the light mediator $\phi$. First we consider constraints on the DM self-interaction coupling $g_\chi$, which allows a DM scattering cross section per mass $\sigma/m\lesssim 1-10~\text{cm}^2/\text{g}$, with the relevant momentum transfer weighted cross section given by~\cite{Tulin:2012wi}
\begin{equation}
\sigma_\text{T} = \frac{4\pi}{m_\phi^2}\beta^2 \log \left(1+\beta^{-1}\right),
\end{equation}
with $\beta = 2\alpha m_\phi/(m_\chi v^2)$ and $v \sim30$ km/sec for typical dwarf galaxies (the expression being valid for $\beta <0.1$ and $m_\chi v/m_\phi\gg 1$). Taking the DM mass to be $\sim m_n$, we get $g_\chi\lesssim 4\times 10^{-4}$ with a mild dependence on the mediator mass $m_\phi$. 
{For a $g_\chi$ value slightly smaller  than $4\times 10^{-4}$, the DM self-interactions fall in the range that alleviate the problems of CCDM. At the same time, the constraints on light particles coupling to the neutron allow us to find parameter space satisfying our NS stability condition.} We can for example choose $m_\phi\sim 0.1$~eV and $g_n\sim -10^{-14}$, corresponding to $\mu\sim-0.4~\text{eV}$. Such a value of $g_n$ is allowed by the strict constraints set on arguments of rapid red giant star cooling, (see e.g., Fig.~3 in Ref.~\cite{Heeck:2014zfa}). These values correspond to $z\sim 50$~MeV, which is below the value  71 MeV that we found sufficient to stabilize heavy neutron stars. {We should stress here that close to the surface of the NS where $n_F\rightarrow 0$, one cannot exclude the presence of DM, simply because at low densities $\mu_n \simeq m_n$ while $\mu_{\chi}\simeq m_{\chi}$. Since $m_n>m_{\chi}$, close to the surface there could be neutrons converting to DM. This can be seen at low densities in our Fig.~1. However, this does not change our conclusions. Firstly a small amount of DM at low densities close to the surface does not change the overall stability of the star which depends on the EoS at the center. Secondly, at the NS crust there are heavy nuclei  and the EoS we have been using are not accurate. }

The light mediator $\phi$ could potentially create problems during BBN and CMB since it could contribute to the effective number of relativistic degrees of freedom. To be on the safe side, we require that $\phi$ decays before the start of BBN (i.e., when the Universe was 1 sec old) and before the neutrino decoupling era (which is roughly the same as the start of BBN), in order not to disturb the abundances of light elements. This can be achieved by decaying to active or sterile neutrinos as e.g. in~\cite{Kouvaris:2014uoa}. In the case of sterile neutrinos, the decay can take place via a term $y_N \phi N^cN$ where $N$ is a light sterile neutrino. The requirement that $\phi$ decays before BBN leads to the condition $y_N>2 \times 10^{-7}(0.1\text{eV}/m_{\phi})^{1/2}$. Alternatively, $\phi$ can decay also to active neutrinos 
via an effective operator of the form $\phi (LH)^2/\Lambda^2$ as long as the scale $\Lambda \lesssim 6\times 10^6\sqrt{m_\phi/0.1\text{eV}}$~GeV. This induces an effective coupling between $\phi$ and active neutrinos of the order or larger than $v_{\text{EW}}^2/\Lambda^2=1.7 \times 10^{-9}$. Neutrino dump experiments can in principle set constraints on couplings of neutrinos to other light particles (like $\phi$) (see e.g., \cite{Bauer:2018onh} for a review). Coupling to muon- and tau- neutrinos leads to less severe constraints which are easily satisfied by a coupling of the order of $10^{-9}$ that we need for the decay of $\phi$ before BBN.

\begin{figure}
\includegraphics[width=0.5\textwidth]{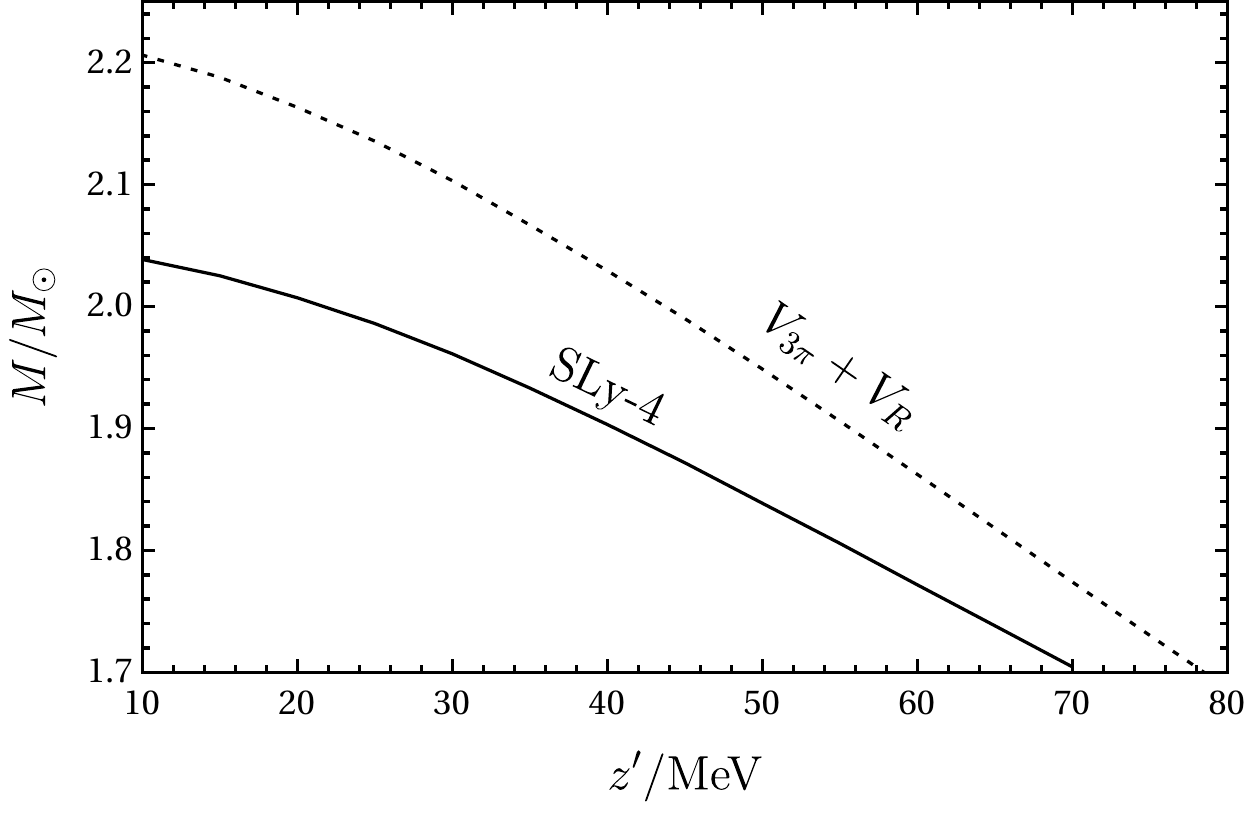}
\caption{The maximum NS mass with DM self-interactions as a function of interaction parameter $z'$. The figure shows two different EoS, the SLy-4 and $V_{3\pi}+V_R$. SLy-4 is relatively soft and $M \geq 2M_\odot$ requires $z'\lesssim 25$~MeV, whereas $V_{3\pi}+V_R$ requires $z'\lesssim 45$~MeV. Both cases assume that the mass of DM is one MeV below that of the neutron.}
\label{Fig: self-int Mch}
\end{figure}

%
%
\paragraph{\bf Self-interactions:} An alternative to DM-baryon cross interactions, although qualitatively different way of accommodating the neutron decays and have NSs with masses above $2M_{\odot}$, is via DM self-interactions. In the case of DM-baryon interactions, the repulsive interaction disfavors the existence of DM inside the star. In the case of DM self-interactions, DM is present but it has a relative stiff EoS because of repulsive self-interactions. As it was confirmed also in~\cite{Cline:2018ami}, in order to support heavy NSs via DM self-interactions, a light vector mediator is needed. If DM is produced thermally, the relic density must be determined by the DM annihilations to these mediators. However, the mediators themselves have to decay to SM particles before the advent of the BBN era. This sets a minimum coupling between the mediator and SM, which is sufficient to deplete most of the DM relic density. Therefore in such a setup, the DM particle that the neutron decays to, cannot account for the whole relic density of DM. This outcome can however be avoided if DM is created with a matter/anti-matter asymmetry. The $t$-channel annihilation rate of DM fermions into light vector mediators is $\sigma_\text{ann} v= g_\chi^4/(16\pi m_\chi^2)(1-m_\phi^2/m_\chi^2)^{1/2}$~\cite{Tulin:2013teo}, which can be much larger than the typical weak scale annihilation rate for couplings larger than $g_\chi \sim 2\times 10^{-2}$. In this case the whole population of anti-DM is depleted via annihilations, leaving only the component in excess. However, large values of the DM coupling $g_\chi$ are forbidden due to constraints on DM self-interactions from the bullet cluster
and other systems. We found that  a coupling $g_\chi = m_\phi/(45~\text{MeV})$ satisfies the small structure constraints $\sigma/m< 10$~cm$^2$/g for $m_\phi\gtrsim 1$~MeV{, and are in fact in the right range to allieviate the CCDM small structure problems}.  For such a mediator mass the decay channel to $e^+e^-$ opens up and it is easy to accommodate the particle's decay before BBN (see Fig.~2 of Ref.~\cite{Cline:2018ami}).Therefore as long as these constraints are satisfied and $g_{\chi}$ is much larger than $2\times 10^{-2}$, asymmetric DM models would be able to account for the whole DM relic density. However, as can be seen from Fig.~\ref{Fig: self-int Mch}, in order to have NSs with mass over $2M_{\odot}$, $z' = m_\phi/g_\chi$ must be smaller than a given value. For example the  SLy-4 EoS requires $z'\lesssim 25$~MeV, whereas $V_{3\pi}+V_R$ requires $z' \lesssim 45$~MeV. Given the conditions we stated above, i.e., having a $g_{\chi}\gg2\times 10^{-2}$ and satisfying the DM self-interactions constraints, one can see that
$2M_{\odot}$ NSs with allowed DM self-interactions and a SLy-4 EoS cannot be achieved, whereas it will be marginally possible for $V_{3\pi}+V_R$. For an ultra hard EoS where $z'$ can even be as large as $60$~MeV~\cite{Cline:2018ami}, DM self-interactions can accommodate easier the maximum mass of $2M_{\odot}$. However, one can see that the DM-baryon cross interactions we presented in this paper provide a better solution to the problem because there is neither a need for strong DM self-interactions which are constrained by observations nor an asymmetric DM production mechanism.

\paragraph{\bf Conclusions:}
In case where the neutron decay discrepancies are due to partial neutron decays to DM, we have shown that a repulsive interaction between DM and neutrons can disfavor the conversion of neutrons to DM inside NSs, thus allowing NSs to be heavier than $2M_{\odot}$, as supported by observations. We propose a microscopic model that can accommodate the DM-baryon cross interactions relying on a light scalar and neutron interactions via a Higgs portal. This is qualitatively different from previous proposals because in our scenario, it is the absence of DM that saves heavy NS from collapsing. In addition, we show that DM self-interactions can also support $2M_{\odot}$ NSs by hardening the EoS. In that case the DM particle can naturally account for the whole relic density as long as it is of asymmetric nature.

  \begin{acknowledgments}
CK is grateful to Mark Alford and Steven Harris for valuable comments on EoS. The work of BG was supported in part by the DOE Grant No. DE-SC0009919. CK is partially funded by the Danish National Research Foundation, grant number DNRF90, and by the Danish Council for Independent Research, grant number DFF 4181-00055.
\end{acknowledgments}

\appendix

\end{document}